%% file: HIN-19-004_temp.tex
\pdfoutput=1
\documentclass[11pt,twoside,a4paper,cmspaper,final,collab]{cms-tdr}
\def\svnVersion{1eafa33-D}\def\svnDate{2022/07/18}\def\cmsCernNoTag{CERN-EP-2022-036}\def\cmsCernDate{\today}\def\cmsMessage{Published in the Journal of High Energy Physics as \href{http://dx.doi.org/10.1007/JHEP05(2023)007}{\doi{10.1007/JHEP05(2023)007}.}}
\begin{document}\cmsNoteHeader{HIN-19-004}

\newcommand{\pp}    {\ensuremath{\text{pp}}\xspace}
\newcommand{\pPb}   {\ensuremath{\Pp\mathrm{Pb}}\xspace}
\newcommand{\PbPb}  {\ensuremath{\text{PbPb}}\xspace}
\newcommand{\noff}  {\ensuremath{N_\text{trk}^\text{offline}}\xspace}
\newcommand{\Vz}    {\ensuremath{\mathrm{V}^\text{0}}\xspace}
\newcommand{\vsub}  {\ensuremath{v_2\{4,\text{Sub}\}}\xspace}

\cmsNoteHeader{HIN-19-004}
\title{Strange hadron collectivity in \texorpdfstring{\pPb}{pPb} and \texorpdfstring{\PbPb}{PbPb} collisions}

\date{\today}

\abstract{
The collective behavior of \PKzS and \PGL/\PAGL strange hadrons is studied by measuring the elliptic azimuthal anisotropy ($v_2$) using the scalar-product and multiparticle correlation methods. Proton-lead (\pPb) collisions at a nucleon-nucleon center-of-mass energy $\sqrtsNN = 8.16\TeV$ and lead-lead (\PbPb) collisions at $\sqrtsNN = 5.02\TeV$ collected by the CMS experiment at the LHC are investigated. Nonflow effects in the pPb collisions are studied by using a subevent cumulant analysis and by excluding events where a jet with transverse momentum greater than 20\GeV is present. The strange hadron $v_2$ values extracted in \pPb collisions via the four- and six-particle correlation method are found to be nearly identical, suggesting the collective behavior. Comparisons of the \pPb and \PbPb results for both strange hadrons and charged particles illustrate how event-by-event flow fluctuations depend on the system size.
}

\hypersetup{
pdfauthor={CMS Collaboration},
pdftitle={Strange hadron collectivity in pPb and PbPb collisions},
pdfsubject={CMS},
pdfkeywords={CMS, heavy ion, scalar product, cumulant, strangeness, V0}}

\maketitle

\section{Introduction}

Strong collective azimuthal correlations between particles have been observed in relativistic nucleus-nucleus collisions 
at the BNL RHIC~\cite{STAR:2002svs,PHENIX:2003qra,PHOBOS:2006dbo} 
and the CERN LHC~\cite{Chatrchyan:2011eka,Chatrchyan:2012wg,Aamodt:2010pa,ATLAS:2012at,Chatrchyan:2012ta,CMS:2013bza}.
These correlations suggest the formation of a strongly interacting quark-gluon plasma (QGP) that exhibits nearly ideal hydrodynamic behavior~\cite{Ollitrault:1992bk,Heinz:2013th,Gale:2013da}.
In recent years, similar long-range collective correlations have been observed for events with high final-state particle multiplicity in both proton-proton (\pp)~\cite{Khachatryan:2010gv,Aad:2015gqa,Khachatryan:2015lva,Khachatryan:2016txc} and proton-lead (\pPb)~\cite{CMS:2012qk,alice:2012qe,Aad:2012gla,Aaij:2015qcq} collisions at the LHC.
These observations have prompted the question as to whether a fluid-like QGP state is also created in lighter hadronic collision systems.
Such long-range correlations have also been  studied at RHIC using
deuteron-gold~\cite{Adare:2014keg,Adamczyk:2015xjc} and helium-3--gold~\cite{Adare:2015ctn} collisions.

Hydrodynamic models provide macroscopic descriptions of a system that is in approximate local thermal equilibrium~\cite{Heinz:2013th}.
The detailed azimuthal correlation structure of emitted particles in these models is
typically characterized by its Fourier coefficients~\cite{Voloshin:1994mz}. In particular,
the second Fourier coefficient, known as elliptic flow ($v_2$),
most directly reflects the medium response to the initial collision geometry, and provides crucial insights into fundamental transport
properties of the QGP medium~\cite{Alver:2010dn,Schenke:2010rr,Qiu:2011hf}.
The properties of the $v_2$ coefficients observed in lighter systems
are consistent with hydrodynamic models of a tiny QGP droplet~\cite{Nagle:2018nvi},
but other explanations have also been suggested for the observed behavior.
A model where gluon saturation in the initial state is assumed has been shown to
capture the main features of the correlation data~\cite{Dusling:2013oia,Dusling:2017dqg}.
Another suggestion is that the observed correlations arise from the anisotropic escape of partons 
which undergo different numbers of scatterings as they traverse the interaction region~\cite{He:2015hfa}.

For a given collision impact parameter, event-by-event fluctuations in the initial-state geometry result in a dispersion of the $v_2$ values.
Both the mean and variance of this distribution can be characterized by cumulants.
The second-order cumulant of strange hadron $v_2$ in \pp~\cite{Khachatryan:2016txc} and \pPb~\cite{Khachatryan:2014jra,Sirunyan:2018toe} collisions has been measured via long-range correlations with a pseudorapidity gap of at least two units.
Here, we present measurements of higher-order cumulants of $v_2$ for the 
strange quark particles \PKzS and \PgL (generically referred to as ``strange \Vz particles''), where the inclusion of the charge-conjugate \PagL is implied.
The results are presented for very high multiplicity \pPb collisions at a nucleon-nucleon center-of-mass energy $\sqrtsNN = 8.16\TeV$ and for lead-lead (\PbPb) collisions at $\sqrtsNN = 5.02\TeV$, including ones with similar multiplicity to the \pPb data.

The $v_2$ coefficients for strange \Vz particles at midrapidity ($\abs{y}<1$) are determined in the transverse momentum (\pt)
range up to 8.5\GeV via multiparticle cumulant and scalar-product (SP) methods~\cite{Luzum:2012da}.
The SP method effectively measures the second-order cumulant of $v_2$ by imposing a pseudorapidity gap.
The multiparticle $Q$-cumulant technique~\cite{Bilandzic:2010jr} is used to determine higher order cumulants involving more than two particles. 
This method has been employed in previous analyses for identified particles in \PbPb collisions by the ALICE~\cite{Zhu:2751439} and CMS~\cite{CMS:2021qqk} collaborations.
In this analysis, the \PbPb results are obtained in an extended range of centrality (5--80\%), 
where centrality is defined in terms of the fraction of the total inelastic nucleus-nucleus cross section, with 0\% denoting the maximum overlap of the colliding nuclei.

Differences in the 4- and 6-particle $v_2$ values observed for high-multiplicity \pPb collisions are examined using a jet veto to suppress jet correlations, which are not related to the overall system collectivity.
The results provide a quantitative measure of the jet contributions to the multiparticle correlation in \pPb collisions as a function of the final-state particle multiplicity.
A subevent analysis of higher-order cumulants is also presented to suppress the effect of short-range correlations in both the \PbPb and \pPb systems.
Tabulated results are provided in the HEPData record for this analysis~\cite{hepdata}.

\section{Experimental setup and data sample}

The CMS detector comprises a number of subsystems.
The silicon tracker, located in the 3.8\unit{T} field of a
superconducting solenoid, consists of 1440 silicon pixel and 15\,148 silicon strip detector modules (Phase-0).
In 2017, an additional layer was added in both the barrel and endcap regions of the pixel detector and the number of silicon pixel modules increased to 1856 (Phase-1)~\cite{CMS-DP-2020-049}.
The silicon tracker measures charged particles within the laboratory pseudorapidity range $\abs{\eta}< 2.5$,
and provides 25--90 (20--75) \mum transverse impact parameter resolution in the Phase-0 (Phase-1) pixel detector and a \pt resolution better than 1.5\% up to 100\GeV~\cite{Chatrchyan:2014fea}.
The electromagnetic (ECAL) and hadron (HCAL) calorimeters are also
located inside the solenoid and cover the range $\abs{\eta} < 3.0$.
The ECAL consists of lead tungstate crystals arranged in a quasi-projective geometry.
The HCAL barrel and endcaps are sampling calorimeters composed of brass and scintillator plates.
Steel and quartz-fiber Cherenkov hadron forward (HF) calorimeters cover the range $2.9 < \abs{\eta} < 5.2$ on either side of the interaction region.
The calorimeter cells are grouped in projective ``towers'' of granularity $(\Delta\eta{\times}\Delta\phi) = 0.087{\times}0.087$ for the central $\eta$ region used in the present jet measurement.
The HF calorimeters are azimuthally subdivided into $20^{\circ}$ modular wedges and further segmented to form $0.175{\times}0.175$ radians $(\Delta\eta{\times}\Delta\phi)$ towers.
The detailed Monte Carlo (MC) simulation of the CMS detector response is based
on \GEANTfour~\cite{GEANT4}. A more detailed description of the CMS detector can be found in Ref.~\cite{JINST}.

The analysis presented in this paper uses approximately $4.27 \times 10^9$ minimum bias (MB) triggered events, corresponding to an integrated luminosity of 0.58\nbinv, from \PbPb collisions collected by the CMS experiment during the 2018 LHC run at $\sqrtsNN = 5.02\TeV$.
The MB events are triggered by requiring signals above the readout threshold of 3\GeV in each of the HF calorimeters~\cite{Khachatryan:2016bia}.
Further selections are applied offline to reject events from background processes (beam-gas interactions and nonhadronic collisions),
as discussed in Ref.~\cite{Khachatryan:2016odn}.
Events are required to have at least one interaction vertex, reconstructed based on two or more tracks,
with a distance of less than 15\cm from the center of the nominal interaction point along the beam axis.
The primary interaction vertex is defined as the one with the highest track multiplicity in the event.
For the tracks used in detecting the primary vertex, 
the shapes of the clusters in the pixel detector have to be
compatible with those expected from particles produced at the primary vertex location.
In the final analysis, the \PbPb collision events are required to have at least two calorimeter towers in each HF detector
with energy deposits of more than 4\GeV per tower. These criteria select $(99\pm2)\%$ of inelastic
hadronic \PbPb collisions. Finding values higher than 100\% reflects the possible presence
of ultra-peripheral (nonhadronic) collisions in the selected event sample.
The event centrality for \PbPb collisions is determined by the transverse energy deposited in the HF calorimeters~\cite{Chatrchyan:2011sx}.

The \pPb data used in this analysis were recorded by CMS in 2016 and corresponds to an integrated luminosity of 186\nbinv~\cite{CMS-PAS-LUM-17-002}.
The beam energies were 6.5\TeV for protons and 2.56\TeV per nucleon for lead nuclei, resulting in $\sqrtsNN = 8.16\TeV$.
A mid-run reversal of the lead and proton beam directions results in negligible differences in the detector-related systematic effects, and the merged results are reported using the convention that the proton-going direction defines positive pseudorapidity.
The nucleon-nucleon center of mass in the \pPb collisions is not at rest with respect to the laboratory frame because of the energy difference between the colliding particles.
Massless particles emitted at $\eta_\text{cm} = 0$ in the nucleon-nucleon center-of-mass frame will be detected at $\eta = -0.465$ (clockwise proton beam)  or $0.465$ (counterclockwise proton beam) in the laboratory frame.
In this paper, an unsubscripted $\eta$ symbol is used to denote the laboratory frame pseudorapidity.
In order to select high-multiplicity \pPb collisions, a dedicated high-multiplicity trigger was implemented using the CMS level-1 (L1)~\cite{Sirunyan:2020zal}
and high-level trigger (HLT) systems~\cite{Khachatryan:2016bia}.
At L1, the \pPb events were triggered by requiring at least
one track with $\pt > 0.4\GeV$ in the pixel tracker
during a \pPb bunch crossing and at least one tower in one of the two HF detectors having an energy above 1\GeV.
In addition, the total number of ECAL and HCAL towers with transverse energies above a threshold of 0.5\GeV is required to exceed 120 and 150 in ECAL and HCAL, respectively.
The events that pass the L1 trigger are subsequently filtered by the HLT.

The track reconstruction that is performed online, as part of the HLT trigger, uses the identical reconstruction algorithm as employed in the offline processing~\cite{Chatrchyan:2014fea}.
For each \pPb event, the vertex reconstructed with the highest number of primary tracks was selected.
The number (multiplicity) of primary tracks (\noff)
with $\abs{\eta}<2.4$, $\pt > 0.4\GeV$, and a distance of closest approach (DCA) to this vertex of 0.4\cm or
less, were determined for each event. 
Because of hardware limits on the data acquisition rate, a fraction of all triggered events were recorded (\ie ``prescaled'').
No prescale was applied on the highest multiplicity events.
In the offline analysis, hadronic collisions are selected by the requirement of a
coincidence of at least one HF calorimeter tower containing more than 3\GeV of
total energy in each of the HF detectors within $3.0<\abs{\eta}<5.2$.
Events are also required to contain at least one reconstructed primary
vertex within 15\cm from the nominal interaction point along the beam axis
and within 0.15\cm transverse to the beam trajectory.
At least two reconstructed tracks are required to be associated
with the primary vertex. Beam-related background is suppressed
by rejecting events for which less than 25\% of all reconstructed tracks
pass the track selection criteria as above.
The entire \pPb data set is divided into classes of reconstructed primary track multiplicity, \noff,
where primary tracks passing the high-purity criteria described in Ref.~\cite{Chatrchyan:2014fea} and with $\abs{\eta}<2.4$ and $\pt >0.4\GeV$ are counted.
The HLT \pt cutoff, which is only applied in determining \noff, is higher than that used for the analysis because of online processing time constraints.
The absence of the time constraints in the offline processing allows extending the \pt coverage down to 0.3\GeV in the analysis.
The multiplicity classification ($120\leq \noff < 150$, $150\leq \noff < 185$, $185\leq \noff < 250$, $\noff \geq 250$) in this analysis is identical to that used in Ref.~\cite{Chatrchyan:2013nka},
where more details are provided, including a table relating \noff to the fraction of MB-triggered events.

Tracks for both \PbPb and \pPb events must pass the high-purity selection criteria described in Ref.~\cite{Chatrchyan:2014fea}.
In addition, a reconstructed track is only considered as a candidate track originating from the primary vertex if the separation along the beam axis 
between the track and the best matching vertex, and the track-vertex impact parameter measured transverse to the beam,
are both less than three times their respective uncertainties. The relative uncertainty
in the \pt measurement is required to be less than 10\%.

The CMS particle-flow (PF) algorithm~\cite{CMS:2017yfk} aims to identify stable particles in an event by combining information from all subdetector systems,
classifying them as electrons, muons, photons, and charged and neutral hadrons.
The PF candidates are then clustered into jets using the anti-\kt sequential recombination algorithm~\cite{Cacciari:2008gp} provided in the \textsc{FastJet} framework~\cite{Cacciari:fastjet1}.
The jets used in this analysis are obtained using a distance parameter $R = 0.4$.
The underlying event contribution to the jet energy in \pPb collisions is subtracted using an iterative procedure described in Refs.~\cite{Kodolova:2007hd,Chatrchyan:2011sx}.

\section{\texorpdfstring{Reconstruction of \PKzS and \PgL}{Reconstruction of K0S and Lambda}}

The reconstruction procedure for \PKzS and \PgL candidates in \PbPb collisions at $\sqrtsNN = 5.02\TeV$ is presented in Ref.~\cite{Khachatryan:2014jra}.
To increase the efficiency for tracks with low momenta and large impact parameters,
both characteristic of the strange-particle decay products, a loose selection of tracks, as discussed in Ref.~\cite{Chatrchyan:2014fea}, was used.
The \PKzS\ and \PgL candidates are reconstructed by combining oppositely charged particles to define a secondary vertex.
The track pair is assumed to originate from $\Pgpp\Pgpm$ in \PKzS reconstruction, and $\Pgpm\Pp\ (\Pgpp\Pap)$  in the \PgL (\PagL) reconstruction.
As the proton (antiproton) carries nearly all of the laboratory-frame momentum for the \PgL (\PagL) decay, the higher-momentum track is assumed to be a proton (antiproton) and the other track a pion.
Each of the two tracks must have hits in at least three layers of the silicon tracker, and transverse and longitudinal significances of the fitted impact parameter with respect to the primary vertex greater than 1.
The significance of a quantity is defined as its value divided by its uncertainty.
The distance of closest approach of the pair of tracks is required to be less than 1.0\cm.

As a result of the long lifetime of the \PKzS and \PgL particles ($c\tau>1\cm$), the significance of the \Vz decay length, which is the three-dimensional (3D)
distance between the primary and \Vz vertices, is required to exceed 2.5.
To remove \PKzS\ candidates misidentified as \PgL\ particles and vice versa,
the \PgL\ (\PKzS) candidate mass, assuming both tracks to be pions (the lower-momentum track to be a pion and the higher-momentum track a proton/antiproton) must differ by more than 20\,(10)\MeV from the nominal \PKzS\,(\PgL) mass value~\cite{ParticleDataGroup:2020ssz}.
To remove photon conversions to an electron-positron pair, the mass of a \PKzS\ or \PgL\ candidate must be higher than 15\MeV assuming both tracks to have the electron mass.
The angle $\theta^{\text{point}}$ between the \Vz momentum vector and the vector connecting the primary and \Vz vertices is required to satisfy $\cos\theta^{\text{point}}>0.999$.
This reduces the contributions of particles from nuclear interactions, random combinations of tracks, and
secondary \PgL\ particles originating from the weak decays of $\Xi$ and $\PgO$ particles.

\begin{figure}[thbp]
\centering
\includegraphics[width=0.49\textwidth]{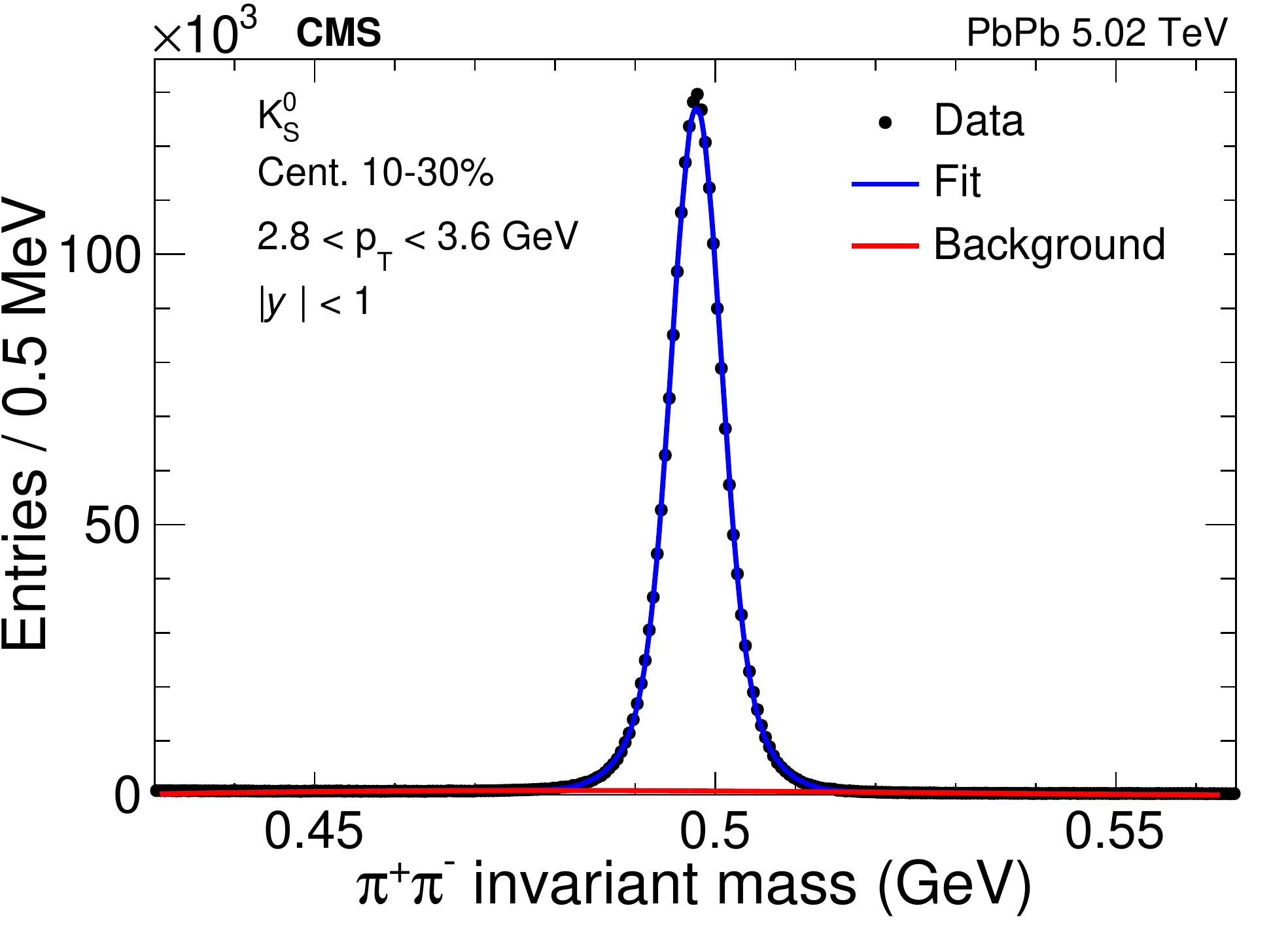}
\includegraphics[width=0.49\textwidth]{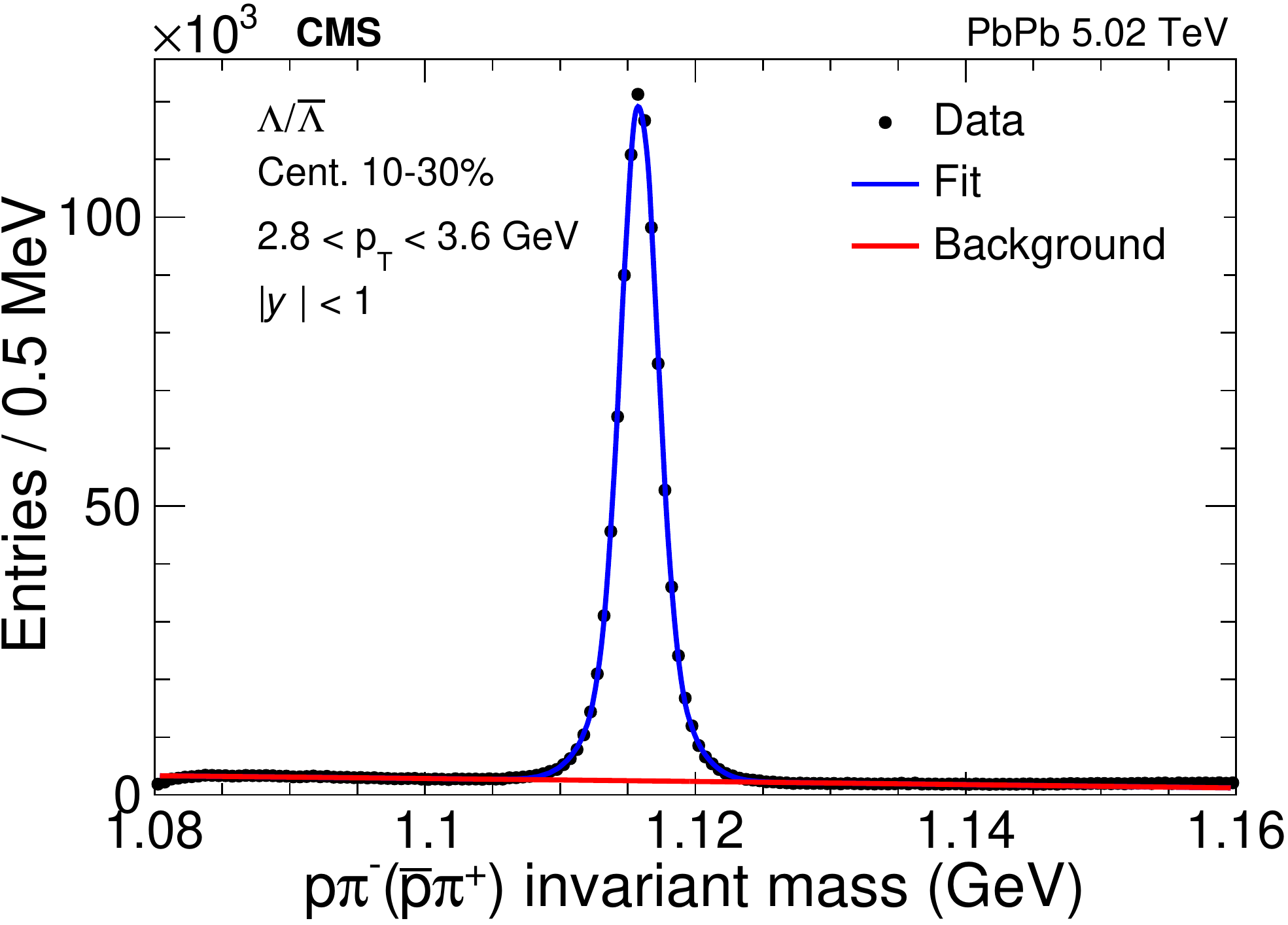}
    \caption{Invariant mass distributions of \PKzS mesons (left) and \PgL baryons (right) candidates within $\abs{y}<1$ and $2.8 < \pt < 3.6\GeV$ in 10--30\% centrality \PbPb collisions at 5.02\TeV. The blue lines show the fitted signal peak. The red lines indicate the fitted background component.}
\label{fig:mass}
\end{figure}

The final selection of strange \Vz candidates for \PbPb collisions is performed via multivariate discriminators to enhance the efficiency and signal-to-background ratio, based on a boosted decision tree (BDT) method from the TMVA package~\cite{Hocker:2007ht}.
The selection is optimized separately for \PKzS and \PgL candidates.
The discriminating variables employed include: the collision centrality,
the \Vz candidate \pt, rapidity, DCA, 3D decay length, $\cos\theta^{\text{point}}$, the fitted 3D vertex $\chi^{2}$, the 3D fitted decay vertex significance,
the \Vz daughter track \pt, uncertainty in \pt, $\eta$, the number of hits in the silicon tracker, the number of pixel detector layers with hits, and the transverse and longitudinal DCA significances of the daughter tracks.
The BDT training is performed employing the \HYDJET 1.8 event generator~\cite{Lokhtin:2005px} with full detector simulation of both the signal (strange \Vz) and random combination of hadrons as background (combinatorial) candidates.
The total average efficiency of detecting \PKzS (\PgL) using the BDT method is 1--3 (1--2)\% from most central to peripheral \PbPb collisions.
Figure \ref{fig:mass} shows examples of the invariant \PKzS and \PgL mass distributions in the \pt interval 2.8--3.6\GeV for 10--30\% centrality in \PbPb collisions.
The \PKzS mass window is taken between 0.430 and 0.565\GeV and the \PgL mass window is taken between 1.08 and 1.16\GeV.
The true \Vz signal peak is well described by a double Gaussian function with a common mean, while the background is modeled by a fourth-order polynomial function fitted over the entire mass range.
The mass window of $\pm2\sigma$ wide around the center of the peak is defined as the ``peak region,'' where $\sigma$ represents the standard deviation of the sum of the two Gaussian distributions.
To estimate the contribution of background candidates in the peak region to the correlation measurement, 
a ``sideband region'' is chosen that includes \Vz candidates from outside the $\pm3\sigma$ mass range around the \Vz mass to the boundaries of the mass distribution windows.

The reconstruction and cut-based selection procedures for the strange \Vz candidates in \pPb collisions are
identical to those in Refs.~\cite{Khachatryan:2014jra,Khachatryan:2016yru}.
Pairs of oppositely charged particle tracks that are detached from the primary vertex (\ie, having large significance of the impact parameters) are selected to determine if they point to a common secondary vertex from the decay of a strange \Vz candidate.
The difference in the cut-based and multivariate selections was found to be significant only for more central ($<30$\% centrality) \PbPb events with \noff values well above those in the \pPb events studied in this analysis and, hence, only a cut-based selection was used for the \pPb events.

\section{Analysis techniques}

The scalar-product~\cite{Luzum:2012da} and multiparticle $Q$-cumulant~\cite{Bilandzic:2010jr,Bilandzic:2013kga} methods are used to study the elliptic flow $v_2$ of charged hadrons
within $\abs{\eta}<1.0$ and \Vz particles within rapidity $\abs{y}<1.0$.
The analysis is similar to that discussed in Refs.~\cite{Sirunyan:2017igb,Sirunyan:2017pan}.
The SP method uses the HF detectors to determine Q-vectors defined by reference flow particles (RFP)
that have a large pseudorapidity gap ($\Delta\eta$) from the analyzed charged hadrons or strange \Vz  particles of interest (POI).
Two Q-vectors are calculated using the HF detectors in either the p-going (p-SP) or Pb-going (Pb-SP) direction.
This method results in the same $v_2$ values as a two-particle correlation analysis with a large $\Delta\eta$ gap.
The gap, here  with $\abs{\Delta\eta} > 2$,  reduces the contributions of nonflow short-range correlations, such as jet fragmentation, the correlation of back-to-back dijets, and resonance decay.
The $Q$-cumulant method has been used to measure  4-, 6-, and 8-particle correlations
in previous CMS publications~\cite{Sirunyan:2019pbr,Sirunyan:2017igb}, with the multiparticle correlations being inherently less sensitive to the few-particle correlations
such as those arising from jet fragmentation and back-to-back dijet correlations.
Differences in the higher-order cumulants are also sensitive to initial-state effects that can lead to a non-Gaussian distribution of flow vectors.
The $m$-particle cumulant method correlates each POI (strange \Vz or charged particles) with $m-1$ RFPs (charged hadrons).
In this analysis, the RFPs for the cumulant method are charged hadrons within $\abs{\eta}<2.4$ and with $0.5 < \pt < 5.0\GeV$.

For the strange \Vz particles, the $v_2$ values are first extracted from the peak region, which contains small contributions from the combinatorial background, and sideband regions, denoted as $v_2^{\text{obs}}$ and $v_2^{\text{bkg}}$, respectively.
The $v_2$ signal of the true \Vz particles is denoted by $v_2^{\text{sig}}$ and is obtained by
\begin{linenomath}
\begin{equation}
    v_2^{\text{sig}} = \frac{v_2^{\text{obs}}-(1-f^{\text{sig}}) v_2^{\text{bkg}}}{f^{\text{sig}}},
\end{equation}
\end{linenomath}
where $f^{\text{sig}}$ represents the signal yield fraction in the peak region determined by the fit to the mass distributions.
The $v_2^{\text{sig}}$ and $v_2^{\text{bkg}}$ values are assumed to be independent of each other.

Nonflow behavior is more obvious in the small collision system because of the relatively low multiplicities involved.
The jet correlations reduce the multiparticle cumulant $v_2\{m\}$ ($m =$ 4, 6, 8) values~\cite{Borghini:2001vi}.
While the jet correlations increase the two-particle correlation $v_2\{\text{SP}\}$ results, 
the large $\Delta\eta$ applied in this analysis minimizes this effect.
To investigate the relative importance of the jet correlations in \pPb collisions, 
a jet veto study is performed by rejecting events in the cumulant analyses with at least one jet with $\pt > 20\GeV$ and within $\abs{\eta} < 2.0$.

In a standard multiparticle $Q$-cumulant analysis, the POI and RFP overlap in $\eta$.
In order to suppress nonflow effects from the short-range correlations in the multiparticle cumulant measurements, a separate analysis is performed where a pseudorapidity gap is imposed between the POI and the RFP~\cite{Jia:2017hbm}.
To implement this gap, the subevent method is used, where the POI and the RFP are selected in nonoverlapping $\eta$ ranges.
In this analysis, the POI is selected to have $0<y<1$ ($-1<y<0$) for \Vz or $0<\eta<1$ ($-1<\eta<0$) for charged hadrons, while the RFP is selected for charged hadrons with $-2.4 < \eta < 0$ ($0 < \eta < 2.4$).
The large underlying event contributions in \PbPb data make it both impractical and unnecessary to apply a jet veto.

\section{Systematic uncertainties}

The sources of systematic uncertainties for the \Vz analysis in \PbPb collisions include reconstruction requirements (BDT selections) and the primary vertex position.
The systematic uncertainties for the charged hadrons in \PbPb collisions are also taken to include the primary vertex positions and track quality requirements similar to Ref.~\cite{Sirunyan:2017igb}.
The uncertainties are quoted  relative to  the $v_2$ values. No centrality or \pt dependence is found for the \PbPb collision results.

The BDT cutoffs have been varied up and down by 5\% to account for the systematic uncertainties associated with the deviations between the simulation and experimental results.
A 3\% systematic uncertainty is quoted by comparing to the \Vz BDT nominal cutoffs in \PbPb collisions.
The uncertainties related to the \Vz background mass modeling are investigated by using an exponential or a third-order polynomial fit, 
and the mass spectrum of same-sign hadron pairs as background distribution, instead of the default fourth-order polynomial fit. 
A 1\% systematic is quoted for the effects related to the background modeling.
We vary the signal yield fraction $f^\text{sig}$ up and down by 3\% to account for the BDT training variable differences between the simulation and data.
This results in a 2\% systematic uncertainty in the final results.
The systematic uncertainties in the primary vertex position along the beam direction ($z_{\text{vtx}}$) have been studied by performing the analysis in different $z_{\text{vtx}}$ ranges.
The systematic uncertainties are estimated as 1\% for charged hadrons, 2\% for \PKzS mesons, and 4\% for \PgL baryons.
For the charged-hadron results, the potential systematic uncertainty related to the track selection criteria is investigated by varying the criteria as described in Ref.~\cite{Sirunyan:2017igb}, and a 1\% systematic uncertainty is quoted.

The systematic uncertainties in \pPb collisions are studied for the \Vz reconstruction (2.0\%), effects of multiple collisions at the same bunch crossing (0.5\%), trigger bias (1.0\%), and the reconstruction of primary vertex position (1.0\%).
A total systematic uncertainty of 2.5\% is quoted. More details can be found in Ref.~\cite{Khachatryan:2016yru}.
For the  jet veto results, we estimate the systematic uncertainties related to the PF jet energy scale and jet energy resolution effects.
For the jet energy scale, we take a conservative estimate of 5\% variation of the cutoff value, rejecting events with jet $\pt > 21\GeV$ and compare the results with the nominal 20\GeV threshold.
No significant difference is observed.
The jet energy resolution was parameterized as shown in Ref.~\cite{Sirunyan:2018jju} for \pp collisions.
The effect of jet energy resolution was studied by applying an additional 20\% uncertainty to the jet \pt,
and no significant effect for a majority of multiplicity and \pt ranges on the $v_2$ results was found.

\section{Results}

\begin{figure}[thbp]
\centering
\includegraphics[width=0.98\textwidth]{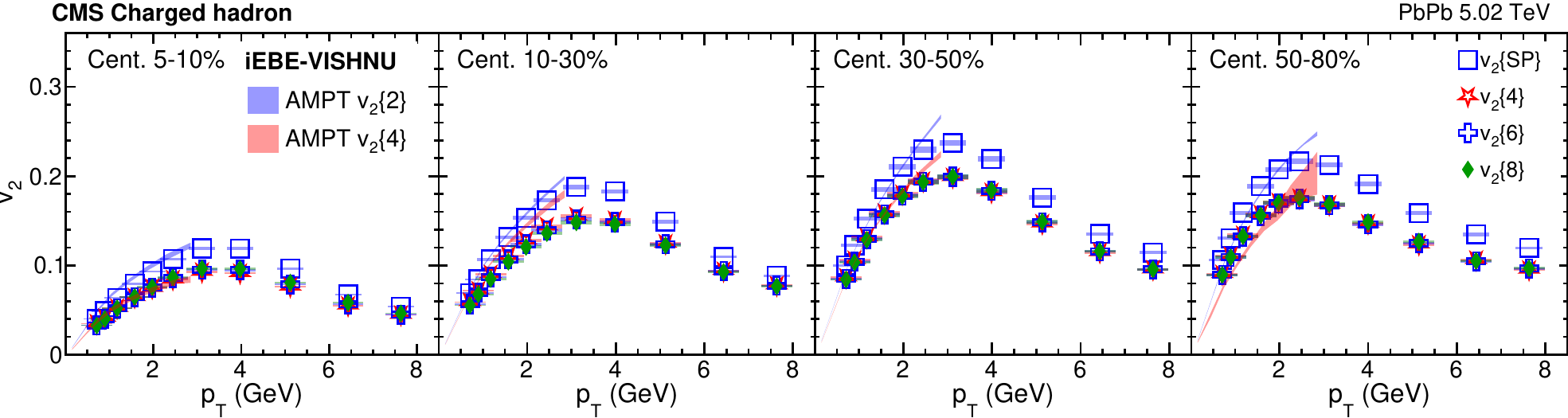}
\includegraphics[width=0.98\textwidth]{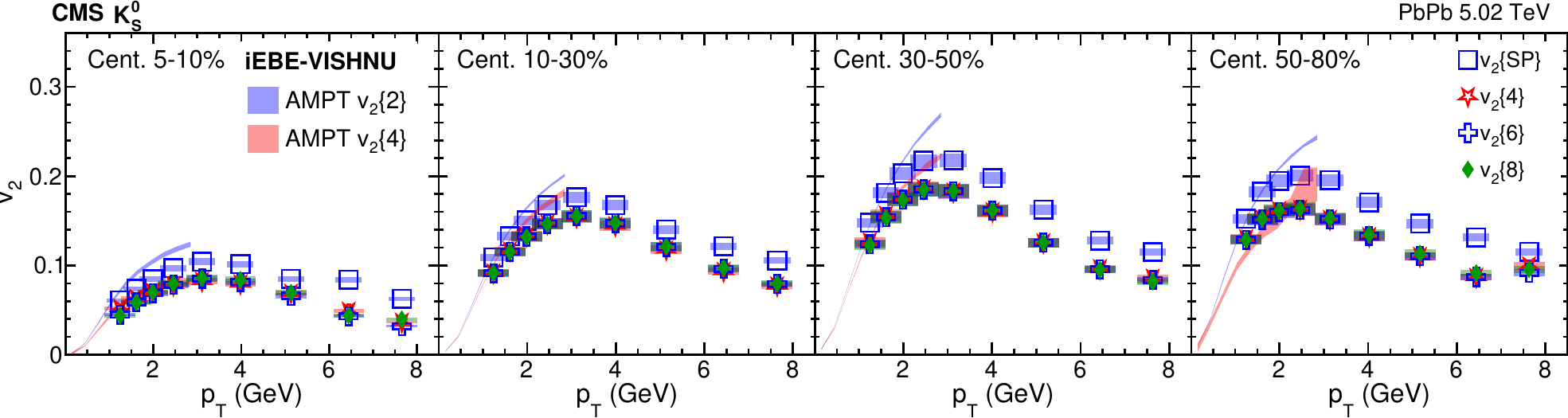}
\includegraphics[width=0.98\textwidth]{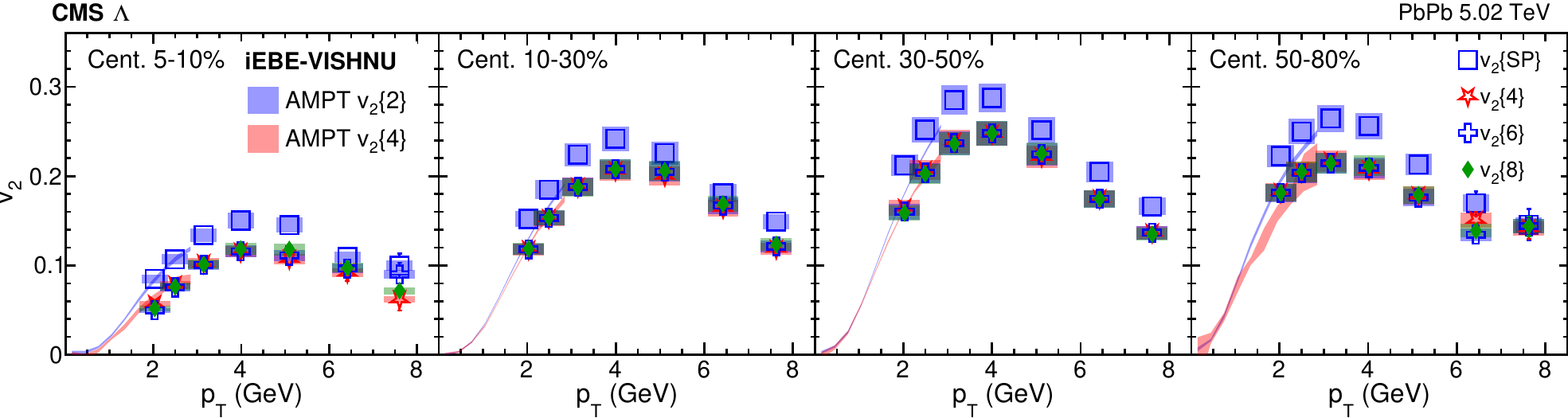}
    \caption{The scalar-product and multiparticle cumulant $v_2$ results of \PbPb collisions at $\sqrtsNN = 5.02\TeV$ for charged hadrons (upper), \PKzS mesons (middle), and \PgL baryons (lower) in different centrality intervals. The shaded bands are hydrodynamic calculations of $v_2\{2\}$ and $v_2\{4\}$ values with AMPT initial conditions. The vertical bars and shaded boxes show the statistical and systematic uncertainties.}
\label{fig:PbPb_v2}
\end{figure}

\begin{figure}[thbp]
\centering
\includegraphics[width=0.98\textwidth]{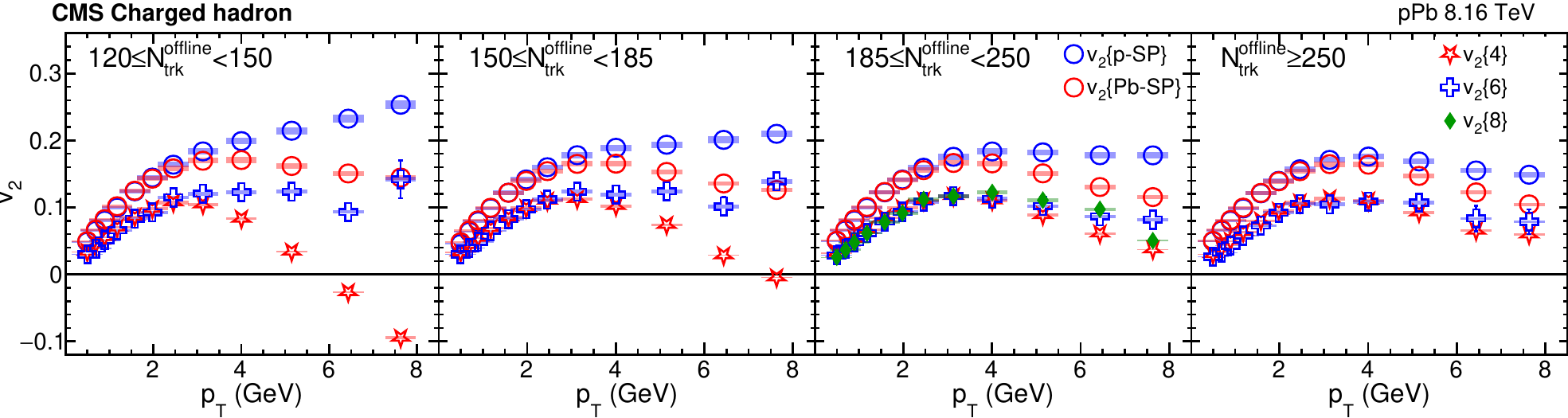}
\includegraphics[width=0.98\textwidth]{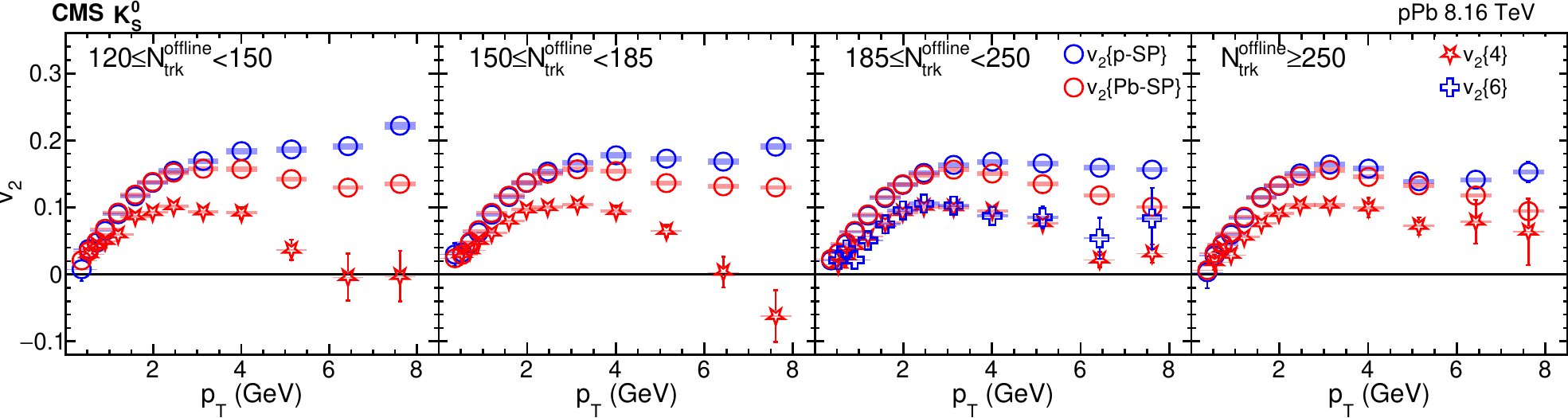}
\includegraphics[width=0.98\textwidth]{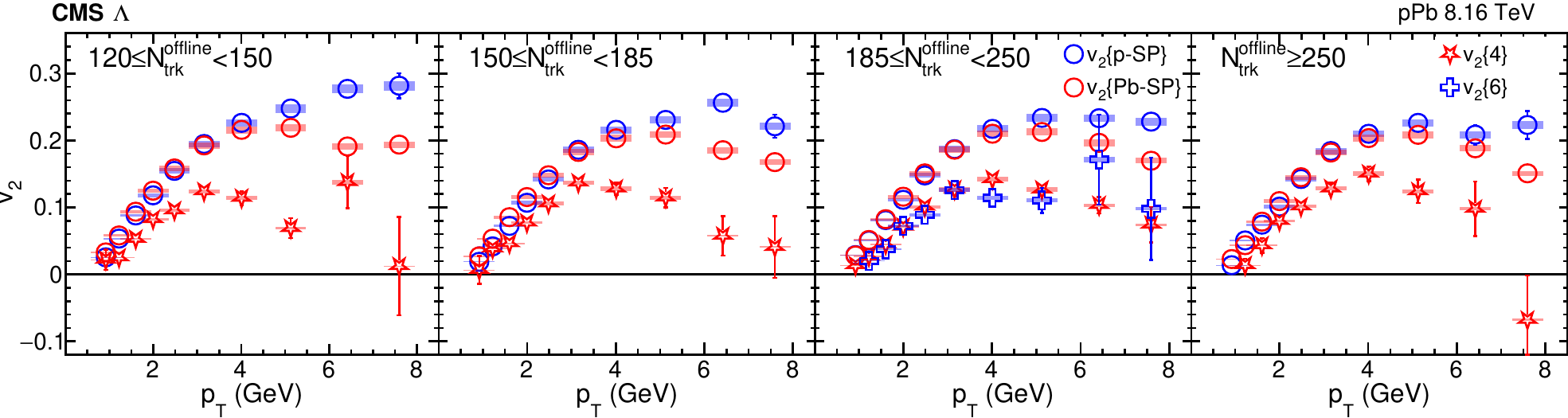}
    \caption{The scalar-product and multiparticle cumulant $v_2$ results of \pPb collisions at $\sqrtsNN = 8.16\TeV$ for charged hadrons (upper), \PKzS mesons (middle), and \PgL\ baryons (lower) in different \noff ranges. The scalar-product results are based on Q-vectors that are determined in either the p-going (p-SP) or Pb-going (Pb-SP) side HF calorimeter. The vertical bars and shaded boxes show the statistical and systematic uncertainties.}
\label{fig:pPb_v2}
\end{figure}

The midrapidity $v_2$ results for charged hadrons within $\abs{\eta}<1.0$, and \PKzS and \PgL particles within $\abs{y}<1.0$ are shown for \PbPb collisions at $\sqrtsNN = 5.02\TeV$ in Fig.~\ref{fig:PbPb_v2}, and for \pPb collisions at $\sqrtsNN = 8.16\TeV$ in Fig.~\ref{fig:pPb_v2}.
The greater values found for $v_2\{\text{SP}\}$ than for the multiparticle flow coefficients at low \pt are consistent with the expectation of the event-by-event fluctuation of the elliptic flow observable.
The event-by-event iEBE-VISHNU model calculations~\cite{Zhao:2017yhj} with A Multi-Phase Transport (AMPT) initial conditions~\cite{Xu:2016hmp}, which combine (2+1)-D viscous hydrodynamics of the QGP expansion with a cascade model of the hadron resonance gas evolution for \PbPb collisions, are compared to the data up to $\pt = 3\GeV$ and found to be qualitatively consistent.
The AMPT model assumes each individual parton has a Gaussian initial energy density profile.
The hydrodynamic $v_2$ values extracted using 2-particle correlation and 4-particle cumulant methods are qualitatively consistent with data in \PbPb collisions.
For the \pPb collisions, the $v_2\{\text{SP}\}$ results with respect to the Pb- and p-going side Q-vectors are presented separately.
For $\pt > 3\GeV$, the p-going Q-vector leads to systematically larger $v_2\{\text{SP}\}$ values, suggesting a larger nonflow contribution.
These $v_2$ results are consistent with those observed in previous \pPb measurements at 5.02\TeV~\cite{Sirunyan:2017igb}.

For the \pPb collisions at 8.16\TeV, a separation of the charged-hadron $v_2\{4\}$ and $v_2\{6\}$ values becomes obvious for lower \noff values and at higher \pt, as shown in Fig.~\ref{fig:pPb_v2}.
The previous \pPb results at 5.02\TeV~\cite{Sirunyan:2017igb} were limited in statistical significance and did not allow for the difference between the  $v_2\{4\}$ and $v_2\{6\}$ values to be clearly established.
The observed difference can be attributed to  jet-related nonflow correlations, which reduce the values of both $v_2\{4\}$ and $v_2\{6\}$.
Indeed, the $v_2\{6\}$ values are expected to be less affected by the jet correlations than the $v_2\{4\}$ values because they involve more particles in the calculation, reducing the influence of few-particle correlations.

\begin{figure}[thbp]
\centering
\includegraphics[width=0.98\textwidth]{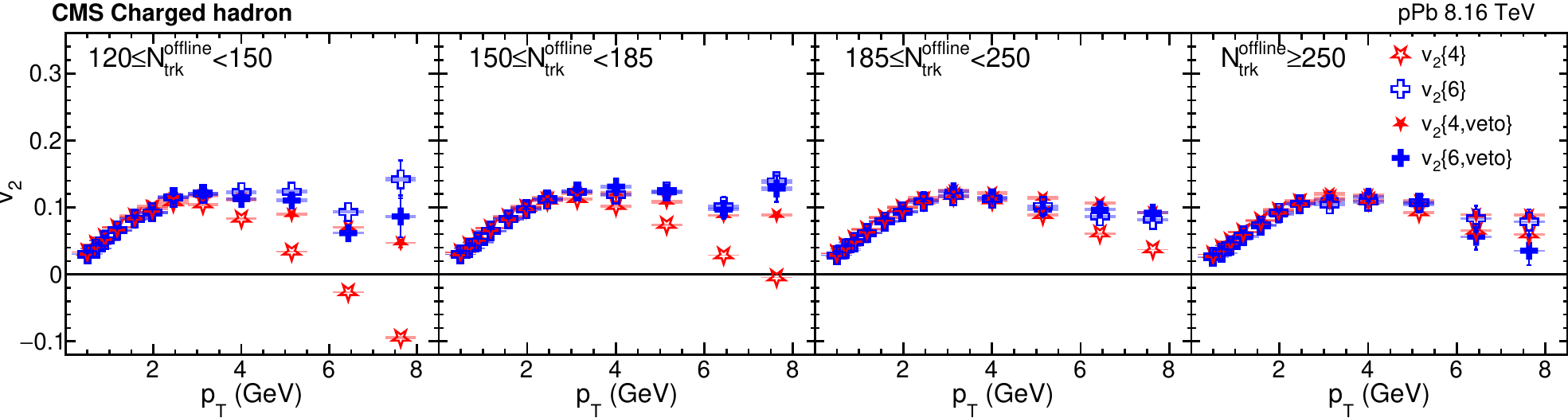}
\includegraphics[width=0.98\textwidth]{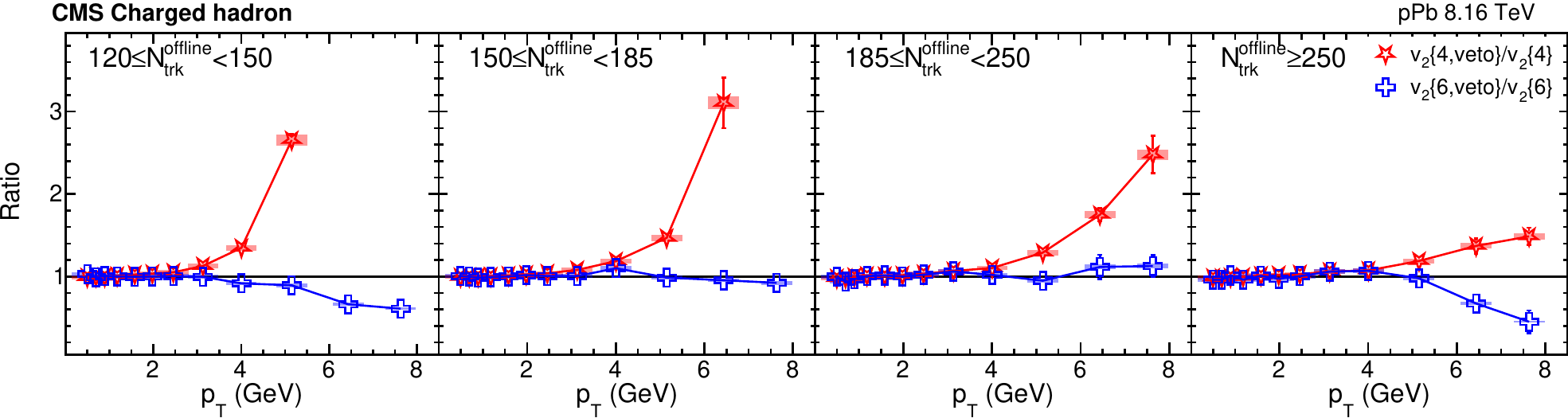}
    \caption{The charged-particle $v_2\{4\}$ and $v_2\{6\}$ distributions for \pPb collisions at $\sqrtsNN = 8.16\TeV$ with and without rejecting jet events (upper) and their ratios (lower) in different \noff ranges. The vertical bars and shaded boxes show the statistical and systematic uncertainties.}
\label{fig:veto}
\end{figure}

To investigate the effect of jet correlations, events with one jet with $\pt > 20\GeV$ are rejected from the \pPb collision sample, leading to the results reported in Fig.~\ref{fig:veto}.
After rejecting the jet events, the 4-particle cumulant $v_2\{4,\text{veto}\}$ value is higher than the inclusive $v_2\{4\}$ value for $\pt > 3\GeV$.
These results indicate that the standard 4-particle cumulant is affected by the jet-related nonflow correlations for all \noff ranges at high \pt, with the influence increasing with decreasing multiplicity as shown in the ratio plots, where the uncertainties are treated as completely correlated.
In contrast, the $v_2\{6\}$ and $v_2\{6,\text{veto}\}$ results are consistent, indicating that the influence of nonflow effects is very small in standard 6- and 8-particle cumulant measurements for the studied multiplicity ranges.
The $v_2\{4,\text{veto}\}$ and $v_2\{6,\text{veto}\}$ values are similar for all \noff ranges, which suggests that, with the jet veto, the jet-related nonflow correlations have been largely removed in the $v_2\{4,\text{veto}\}$ results.

\begin{figure}[thbp]
\centering
\includegraphics[width=0.98\textwidth]{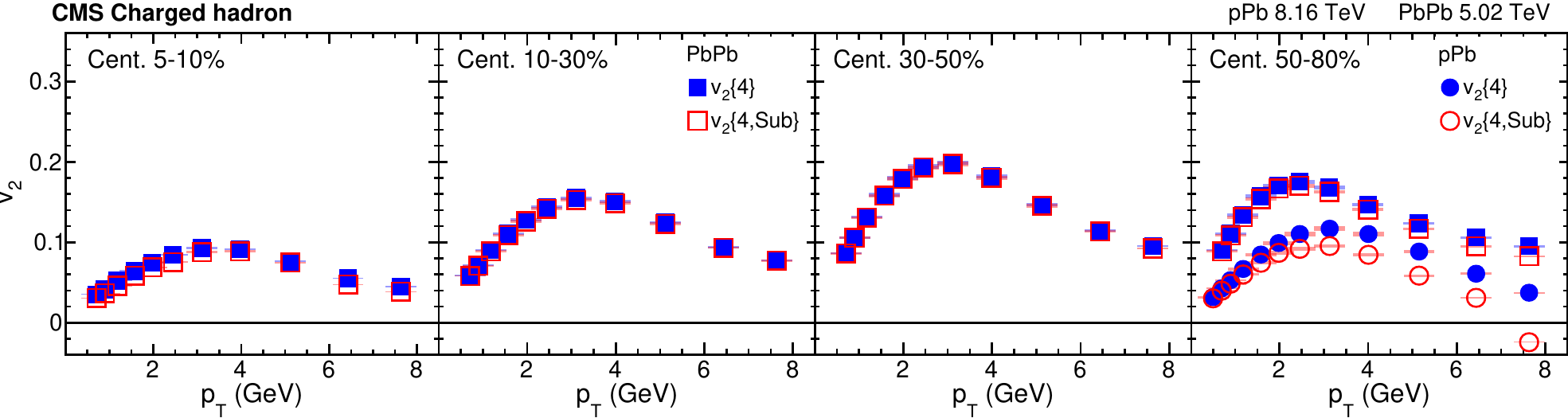}
\includegraphics[width=0.98\textwidth]{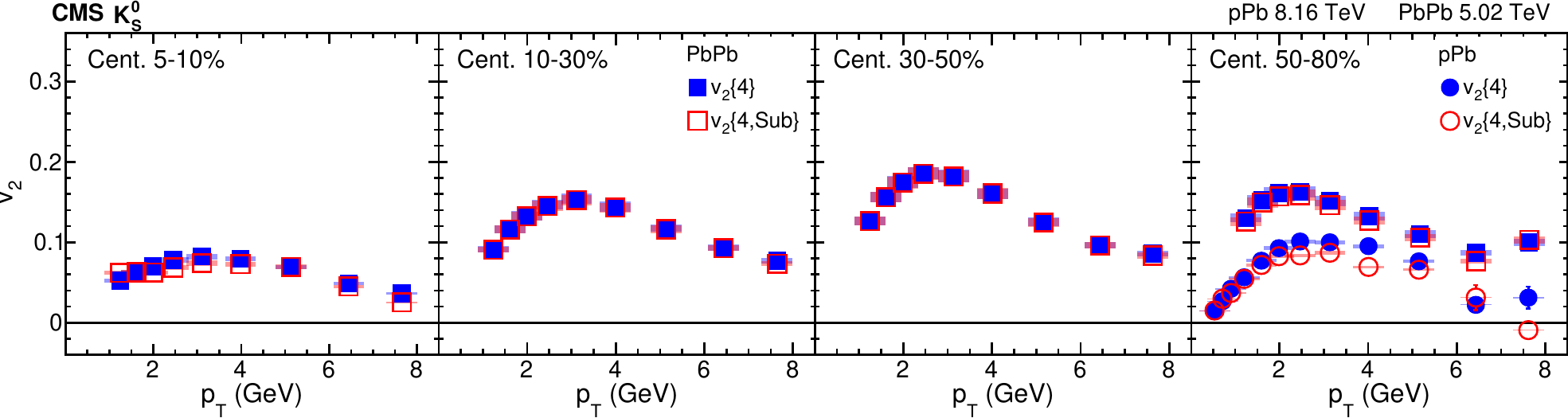}
\includegraphics[width=0.98\textwidth]{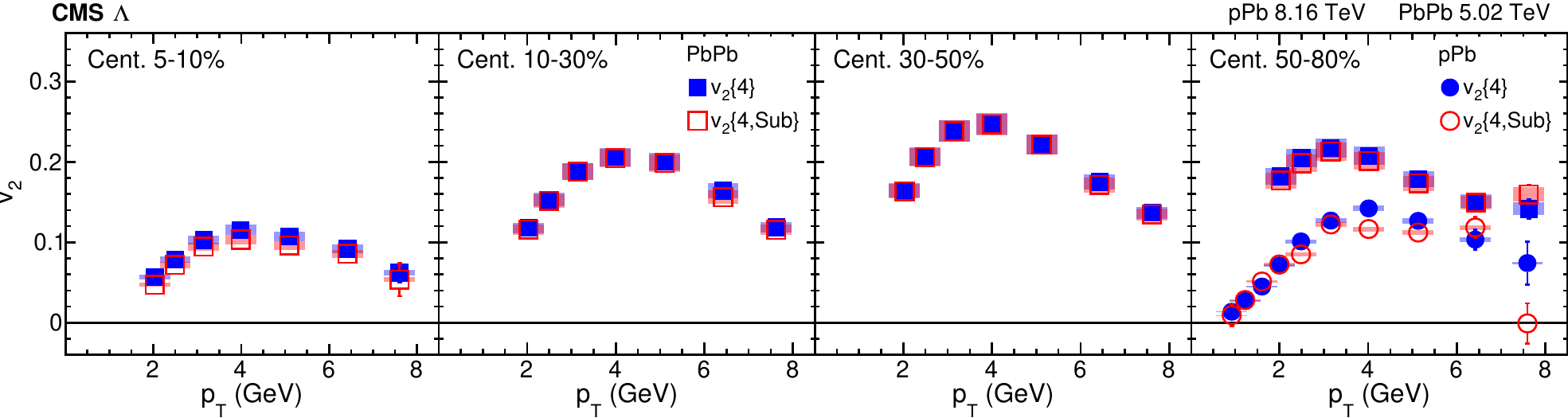}
    \caption{The standard 4-particle $v_2\{4\}$ and subevent 4-particle \vsub values from \PbPb collisions at $\sqrtsNN = 5.02\TeV$ and \pPb collisions at $\sqrtsNN = 8.16\TeV$ for charged hadrons (upper), \PKzS mesons (middle), and \PgL baryons (lower) in different centrality intervals. The vertical bars and shaded boxes show the statistical and systematic uncertainties.}
\label{fig:PbPb_sub}
\end{figure}

\begin{figure}[thbp]
\centering
\includegraphics[width=0.98\textwidth]{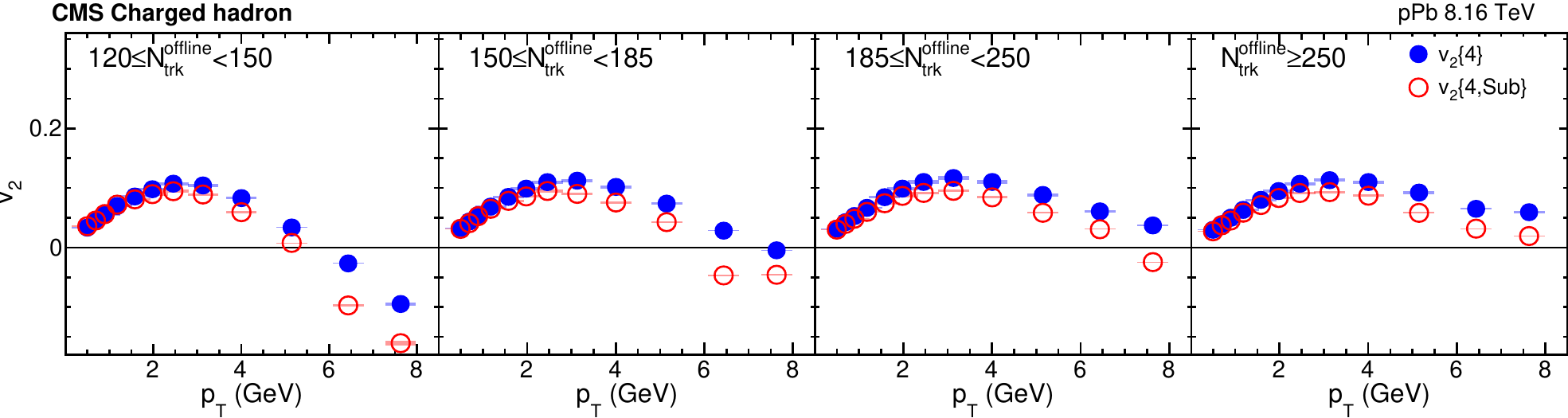}
    \caption{The $v_2\{4\}$ and \vsub values from \pPb collisions at $\sqrtsNN = 8.16\TeV$ for charged hadrons in different \noff ranges. The vertical bars and shaded boxes show the statistical and systematic uncertainties.}
\label{fig:pPb_sub}
\end{figure}

The subevent 4-particle cumulant results \vsub are compared to the standard $v_2\{4\}$ values in Fig.~\ref{fig:PbPb_sub} for charged hadrons and \Vz particles in \PbPb collisions.
Also shown for \pPb collisions are the \Vz standard and subevent 4-particle $v_2$ results for $185 \le \noff < 250$, which corresponds to the multiplicity of about 62\% peripheral \PbPb collisions~\cite{Chatrchyan:2013nka}.
In the mid-central \PbPb collisions (10--50\% centrality), very good agreement is observed between $v_2\{4\}$ and \vsub values,
while a small disagreement is seen in the central (5--10\% centrality) and the peripheral (50--80\% centrality) collisions.
A larger effect is observed in \pPb collisions.
Figure~\ref{fig:pPb_sub} shows the charged-hadron subevent 4-particle results for different \noff ranges in \pPb collisions.

\begin{figure}[thbp]
\centering
\includegraphics[width=0.98\textwidth]{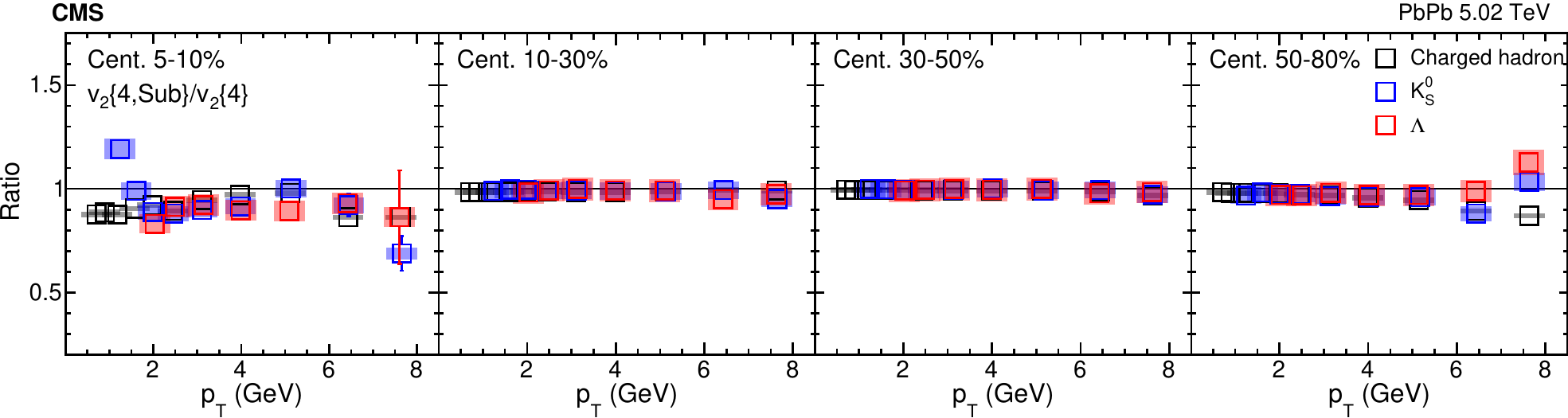}
\includegraphics[width=0.98\textwidth]{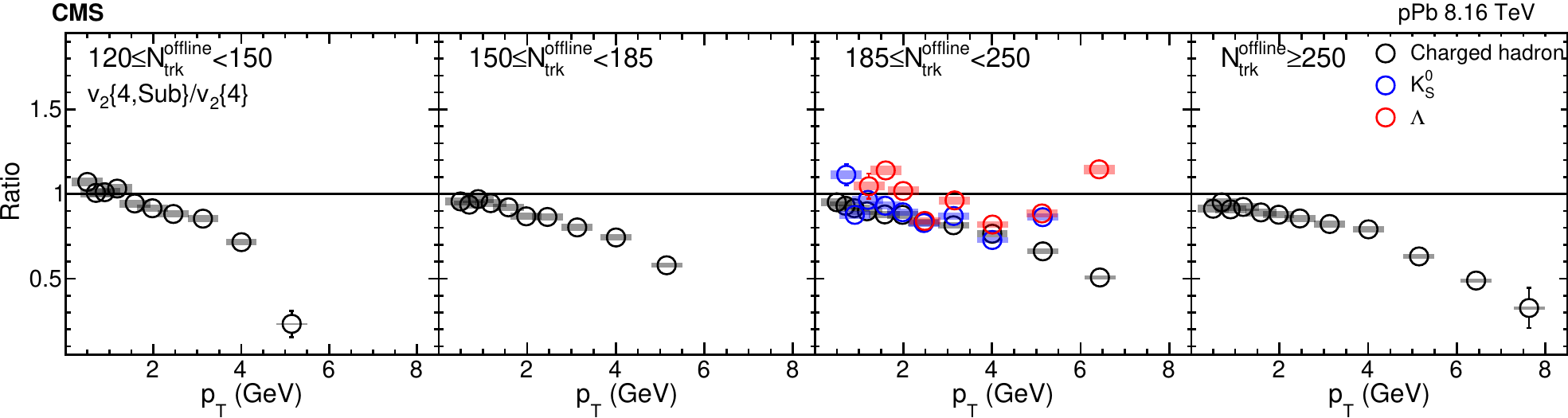}
    \caption{The $\vsub/v_2\{4\}$ ratios for charged hadrons, \PKzS mesons and \PgL baryons, for \PbPb collisions at $\sqrtsNN = 5.02\TeV$ in different centrality intervals (upper) and \pPb collisions at $\sqrtsNN = 8.16\TeV$ in different \noff ranges (lower). The vertical bars and shaded boxes show the statistical and systematic uncertainties.}
\label{fig:subRatio}
\end{figure}

The $\vsub/v_2\{4\}$ ratios for charged hadrons, \PKzS mesons, and \PgL baryons are shown in Fig.~\ref{fig:subRatio} for \pPb and \PbPb collisions.
In \PbPb collisions with centrality in the range of 10 to 80\%, the subevent \vsub values are a few percent smaller than the standard $v_2\{4\}$ values, without having an obvious \pt or particle species dependence.
A larger reduction of up to 10\% is found for the 5--10\% centrality results.
In \pPb collisions, the \vsub values are smaller than the standard $v_2\{4\}$ values by as much as 50\% for higher \pt charged hadrons.
The \pPb identified particle results could only be obtained for the high-statistics $185 \le \noff < 250$ range.
In this range, no strange-quark species dependence is evident.
While the jet veto cumulant $v_2\{4,\text{veto}\}$ values are higher than those observed in the standard method,
an opposite trend is observed by applying the subevent cumulant method.
The \vsub values are consistently lower than those from the standard method, inconsistent with the expectation of nonflow effects being suppressed by imposing $\eta$ gaps among particles.
This might reflect the decorrelation effect of the rapidity gap between the POI and RFP, which is more significant in small collision systems~\cite{Khachatryan:2015oea}.

\begin{figure}[thbp]
\centering
\includegraphics[width=0.98\textwidth]{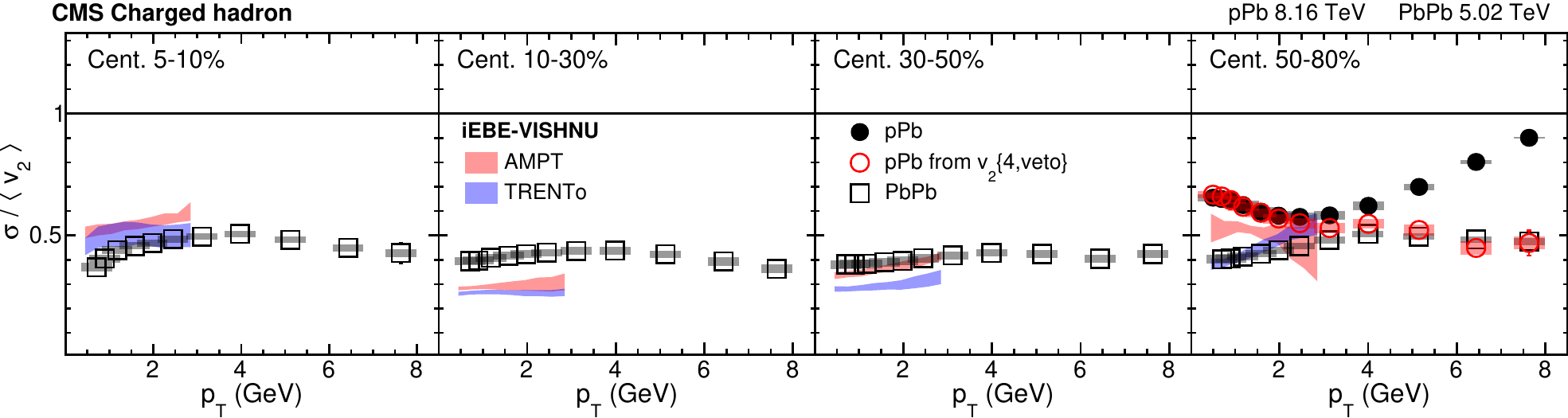}
\includegraphics[width=0.98\textwidth]{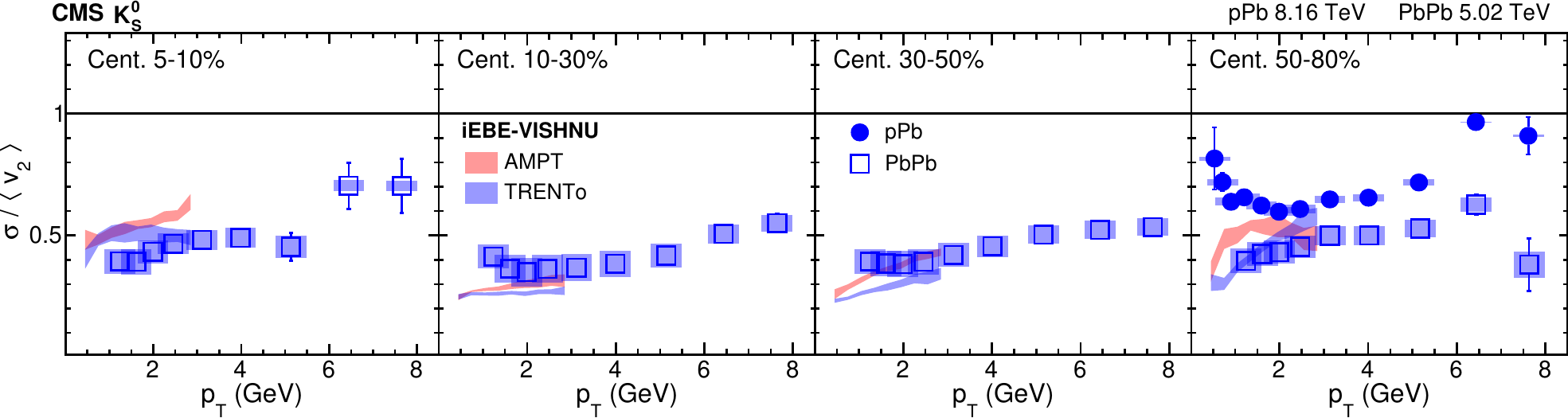}
\includegraphics[width=0.98\textwidth]{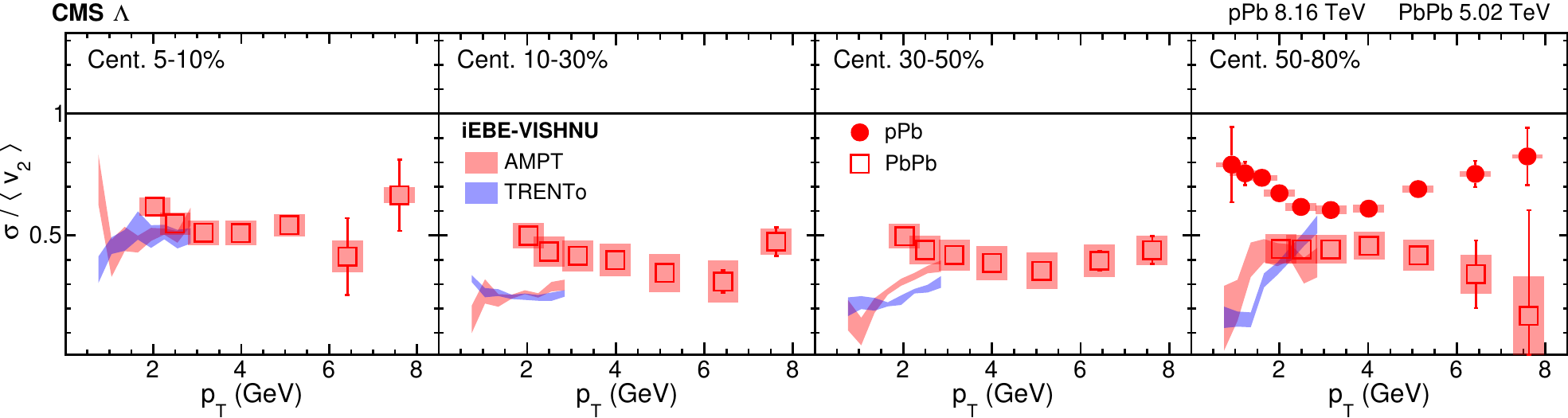}
    \caption{The $v_2$ fluctuation results for \PbPb collisions at $\sqrtsNN = 5.02\TeV$ in different centrality intervals and \pPb collisions at $\sqrtsNN = 8.16\TeV$ with $185 \le \noff < 250$ using $v_2\{\text{Pb-SP}\}$ for charged hadrons (upper), \PKzS mesons (middle), and \PgL baryons (lower). The shaded bands are hydrodynamic calculations of $v_2$ fluctuations with AMPT and TRENTo initial conditions in \PbPb collisions~\cite{Zhao:2017yhj}. The vertical bars and shaded boxes show the statistical and systematic uncertainties.}
\label{fig:v2_fluct}
\end{figure}

\begin{figure}[thbp]
\centering
\includegraphics[width=0.98\textwidth]{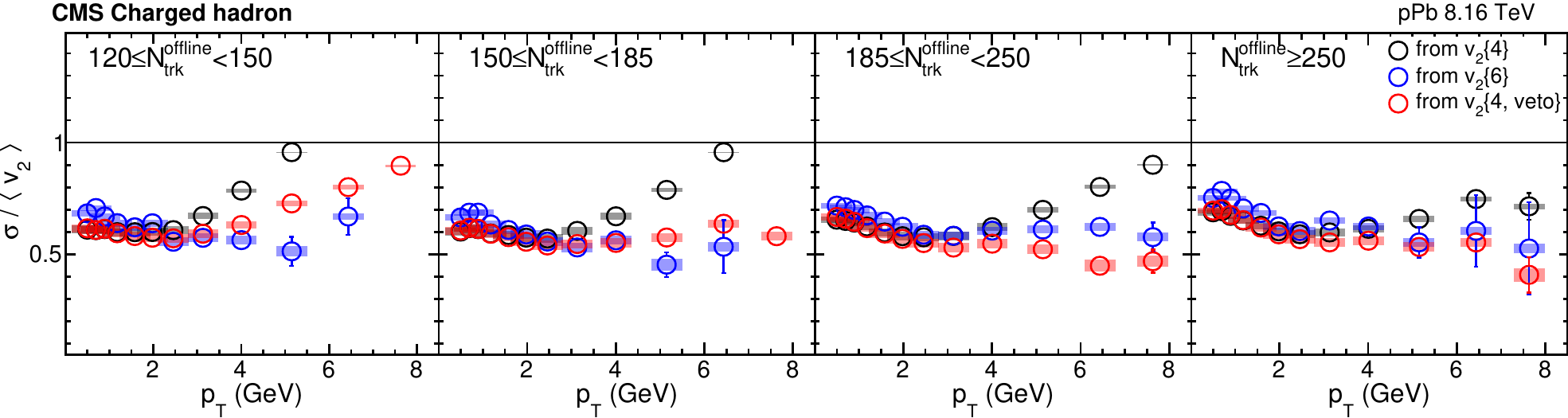}
    \caption{The $v_2$ fluctuation results of \pPb collisions at $\sqrtsNN = 8.16\TeV$ for charged hadrons derived from different multiparticle correlations in different \noff ranges. The vertical bars and shaded boxes show the statistical and systematic uncertainties.}
\label{fig:v2_fluctpPb}
\end{figure}

The difference between the scalar-product $v_2$, as compared to the multiparticle correlation results, is likely to reflect initial-state fluctuation effects.
Event-by-event fluctuations in the location of the participant nucleons can have a large and method-dependent influence on the harmonic coefficients~\cite{Alver:2008zza,Ollitrault:2009ie}.
Expressing the fluctuations in terms of the azimuthal anisotropy in the participant plane $v_2$, the magnitude of the fluctuations is given by $\sigma^2 \equiv \langle v_{2}^2 \rangle - \langle v_{2} \rangle^2$.
At the leading order, the scalar-product and multiparticle cumulant correlations $v_2\{m\}$ ($m=$ 4, 6) are affected differently, with~\cite{Alver:2008zza,Ollitrault:2009ie}
\begin{linenomath}
\begin{equation}
    \begin{aligned}
        v_2\{\text{SP}\}^2 &=& \langle v_2^2 \rangle = \langle v_2 \rangle^2 + \sigma^2,\\
        v_2\{4\}^2 &=& \sqrt{2 \langle v_2^2 \rangle^2 - \langle v_2^4 \rangle} \approx \langle v_2 \rangle^2 - \sigma^2,\\
        v_2\{6\}^2 &=& \sqrt{(\langle v_2^6 \rangle - 9 \langle v_2^4 \rangle \langle v_2^2 \rangle + 12 \langle v_2^2 \rangle^3)/4} \approx \langle v_2 \rangle^2 - \sigma^2.
    \end{aligned}
\end{equation}
\end{linenomath}
The fluctuation ratio $\sigma/\langle v_2 \rangle$ can then be calculated from $v_2\{4\}$, $v_2\{6\}$, and $v_2\{4,\text{veto}\}$ coefficients as
\begin{linenomath}
\begin{equation}
    \begin{aligned}
        \frac{\sigma}{\langle v_2 \rangle} &=& \sqrt{\frac{ v_2\{\text{SP}\}^2 - v_2\{4\}^2 }{ v_2\{\text{SP}\}^2 + v_2\{4\}^2 }}\\
                                    &=&\sqrt{\frac{ v_2\{\text{SP}\}^2 - v_2\{6\}^2 }{ v_2\{\text{SP}\}^2 + v_2\{6\}^2 }}\\
                                    &=&\sqrt{\frac{ v_2\{\text{SP}\}^2 - v_2\{4,\text{veto}\}^2 }{ v_2\{\text{SP}\}^2 + v_2\{4,\text{veto}\}^2 }}.
    \end{aligned}
\end{equation}
\end{linenomath}
The fluctuation ratios of charged hadrons, \PKzS mesons, and \PgL baryons are shown in Fig.~\ref{fig:v2_fluct} for \PbPb in different centrality intervals and compared to those in \pPb collisions with $185 \le \noff < 250$.
The Pb-going side Q-vector $v_2\{\text{Pb-SP}\}$ is used for \pPb collisions.
The fluctuations in \pPb collisions are systematically higher than those in \PbPb collisions for identified particles and charged hadrons.
No clear particle species dependence is observed for either \PbPb or \pPb collisions, suggesting such fluctuations arise from the fluctuations of initial-state geometry.
In \PbPb collisions, the fluctuation results for both strange \Vz particles and charged hadrons are relatively flat over the measured \pt ranges.
In \pPb collisions, the fluctuations are lowest with $\pt\sim3\GeV$, and approach  unity at both low and high \pt.
The iEBE-VISHNU hydrodynamic calculations for \PbPb collisions are also shown with the AMPT and TRENTo initial conditions~\cite{Moreland:2014oya}.
The TRENTo model is optimized to explain experimental results by adjusting a parameterization of the initial entropy density.
Qualitatively consistent with the data, the model calculations show little dependence on \pt.
In \pPb collisions, the charged hadron flow fluctuations calculated from $v_2\{4\}$, $v_2\{6\}$, and $v_2\{4,\text{veto}\}$ coefficients are compared in Fig.~\ref{fig:v2_fluctpPb}.
The flow fluctuations at high \pt are affected by the jet correlations.
If such jet-related nonflow is suppressed by 6-particle correlations, or by the jet veto method, the high-\pt flow fluctuations of charged hadrons in \pPb collisions become compatible with those in the peripheral \PbPb collisions.

\begin{figure}[thbp]
\centering
\includegraphics[width=0.98\textwidth]{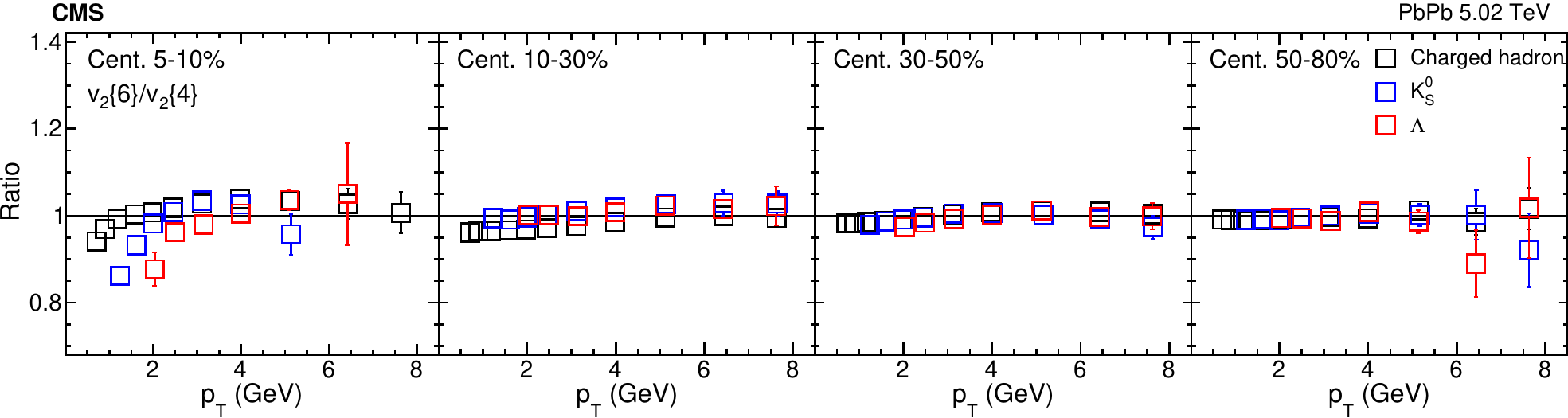}
\includegraphics[width=0.98\textwidth]{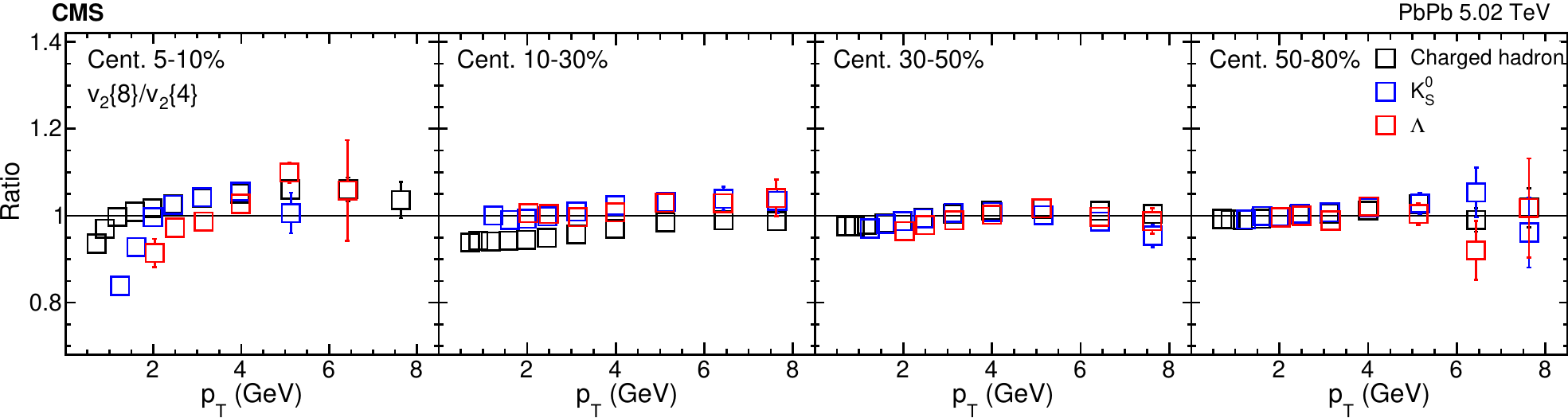}
    \caption{The $v_2\{6\}/v_2\{4\}$ (upper) and $v_2\{8\}/v_2\{4\}$ (lower) ratios in \PbPb collisions at $\sqrtsNN = 5.02\TeV$ for charged hadrons, \PKzS mesons, and \PgL baryons in different centrality intervals. The vertical bars show the statistical uncertainties. The uncertainties are treated as uncorrelated for the ratios.}
\label{fig:PbPb_v2cumu_ratio}
\end{figure}

Assuming Gaussian fluctuations of the flow harmonics, the 4-, 6- and 8-particle correlations should converge, $v_2\{4\} \approx v_2\{6\} \approx v_2\{8\}$.
The differences between $v_2\{6\}$, $v_2\{8\}$, and $v_2\{4\}$ indicate deviation from Gaussian fluctuation behavior for the measured particles in  \PbPb collisions.
The $v_2\{6\}/v_2\{4\}$ and $v_2\{8\}/v_2\{4\}$ ratios are shown in Fig.~\ref{fig:PbPb_v2cumu_ratio} for charged hadrons, \PKzS mesons, and \PgL baryons.
For centrality greater than 30\%, the ratios are independent of \pt and consistent with unity for the studied particle species.

\section{Summary}

The elliptic azimuthal anisotropy $v_2$ values have been measured using the scalar-product and the multiparticle $Q$-cumulant methods for \PKzS mesons, \PgL baryons, and charged hadrons in lead-lead (\PbPb) collisions at $\sqrtsNN = 5.02\TeV$ and high-multiplicity proton-lead (\pPb) collisions at $\sqrtsNN = 8.16\TeV$.
For the first time, multiparticle correlations of identified strange particles have been studied for \pPb collisions.
The hydrodynamic model calculations of scalar-product $v_2\{\text{SP}\}$ and 4-particle $v_2\{4\}$ values for \PKzS mesons, \PgL baryons and charged hadrons using different initial conditions are found to be qualitatively consistent with the observations in \PbPb collisions.
No obvious particle species dependence of the fluctuations in the $v_2$ values is observed for either the \PbPb or \pPb systems, indicating an origin of observed $v_2$ fluctuations from the initial-state geometry.
The flow fluctuations are observed to be larger in \pPb collisions at low transverse momentum (\pt).
However, once jet correlations are removed, the \pPb fluctuations are similar to those observed in peripheral \PbPb collisions at high \pt.

Nonflow effects are studied for the multiparticle cumulant results.
A large difference between the 4- and 6-particle $v_2$ values in \pPb collisions that is not present for \PbPb collisions can be explained by jet-related correlations.
These nonflow correlations are suppressed by rejecting events with at least one jet with $\pt > 20\GeV$.
A subevent cumulant method is also performed to reduce short-range correlation effects in \pPb and \PbPb collisions.
Whereas the jet rejection study results in higher values for $v_2\{4\}$ as compared to the standard method, the subevent method leads to smaller values.
This difference may be attributed to the effect of flow decorrelations on the subevent cumulant method.

\begin{acknowledgments}
We congratulate our colleagues in the CERN accelerator departments for the excellent performance of the LHC and thank the technical and administrative staffs at CERN and at other CMS institutes for their contributions to the success of the CMS effort. In addition, we gratefully acknowledge the computing centers and personnel of the Worldwide LHC Computing Grid and other centers for delivering so effectively the computing infrastructure essential to our analyses. Finally, we acknowledge the enduring support for the construction and operation of the LHC, the CMS detector, and the supporting computing infrastructure provided by the following funding agencies: BMBWF and FWF (Austria); FNRS and FWO (Belgium); CNPq, CAPES, FAPERJ, FAPERGS, and FAPESP (Brazil); MES and BNSF (Bulgaria); CERN; CAS, MoST, and NSFC (China); MINCIENCIAS (Colombia); MSES and CSF (Croatia); RIF (Cyprus); SENESCYT (Ecuador); MoER, ERC PUT and ERDF (Estonia); Academy of Finland, MEC, and HIP (Finland); CEA and CNRS/IN2P3 (France); BMBF, DFG, and HGF (Germany); GSRI (Greece); NKFIH (Hungary); DAE and DST (India); IPM (Iran); SFI (Ireland); INFN (Italy); MSIP and NRF (Republic of Korea); MES (Latvia); LAS (Lithuania); MOE and UM (Malaysia); BUAP, CINVESTAV, CONACYT, LNS, SEP, and UASLP-FAI (Mexico); MOS (Montenegro); MBIE (New Zealand); PAEC (Pakistan); MES and NSC (Poland); FCT (Portugal); MESTD (Serbia); MCIN/AEI and PCTI (Spain); MOSTR (Sri Lanka); Swiss Funding Agencies (Switzerland); MST (Taipei); MHESI and NSTDA (Thailand); TUBITAK and TENMAK (Turkey); NASU (Ukraine); STFC (United Kingdom); DOE and NSF (USA).

\hyphenation{Rachada-pisek} Individuals have received support from the Marie-Curie program and the European Research Council and Horizon 2020 Grant, contract Nos.\ 675440, 724704, 752730, 758316, 765710, 824093, 884104, and COST Action CA16108 (European Union); the Leventis Foundation; the Alfred P.\ Sloan Foundation; the Alexander von Humboldt Foundation; the Belgian Federal Science Policy Office; the Fonds pour la Formation \`a la Recherche dans l'Industrie et dans l'Agriculture (FRIA-Belgium); the Agentschap voor Innovatie door Wetenschap en Technologie (IWT-Belgium); the F.R.S.-FNRS and FWO (Belgium) under the ``Excellence of Science -- EOS" -- be.h project n.\ 30820817; the Beijing Municipal Science \& Technology Commission, No. Z191100007219010; the Ministry of Education, Youth and Sports (MEYS) of the Czech Republic; the Hellenic Foundation for Research and Innovation (HFRI), Project Number 2288 (Greece); the Deutsche Forschungsgemeinschaft (DFG), under Germany's Excellence Strategy -- EXC 2121 ``Quantum Universe" -- 390833306, and under project number 400140256 - GRK2497; the Hungarian Academy of Sciences, the New National Excellence Program - \'UNKP, the NKFIH research grants K 124845, K 124850, K 128713, K 128786, K 129058, K 131991, K 133046, K 138136, K 143460, K 143477, 2020-2.2.1-ED-2021-00181, and TKP2021-NKTA-64 (Hungary); the Council of Science and Industrial Research, India; the Latvian Council of Science; the Ministry of Education and Science, project no. 2022/WK/14, and the National Science Center, contracts Opus 2021/41/B/ST2/01369 and 2021/43/B/ST2/01552 (Poland); the Funda\c{c}\~ao para a Ci\^encia e a Tecnologia, grant CEECIND/01334/2018 (Portugal); the National Priorities Research Program by Qatar National Research Fund; MCIN/AEI/10.13039/501100011033, ERDF ``a way of making Europe", and the Programa Estatal de Fomento de la Investigaci{\'o}n Cient{\'i}fica y T{\'e}cnica de Excelencia Mar\'{\i}a de Maeztu, grant MDM-2017-0765 and Programa Severo Ochoa del Principado de Asturias (Spain); the Chulalongkorn Academic into Its 2nd Century Project Advancement Project, and the National Science, Research and Innovation Fund via the Program Management Unit for Human Resources \& Institutional Development, Research and Innovation, grant B05F650021 (Thailand); the Kavli Foundation; the Nvidia Corporation; the SuperMicro Corporation; the Welch Foundation, contract C-1845; and the Weston Havens Foundation (USA).
\end{acknowledgments}
\bibliography{auto_generated}

\cleardoublepage \appendix\section{The CMS Collaboration \label{app:collab}}\begin{sloppypar}\hyphenpenalty=5000\widowpenalty=500\clubpenalty=5000\input{HIN-19-004-public-authorlist.tex}\end{sloppypar}
\end{document}

%% file: HIN-19-004-public-authorlist.tex
\cmsinstitute{Yerevan Physics Institute, Yerevan, Armenia}
{\tolerance=6000
A.~Tumasyan\cmsAuthorMark{1}\cmsorcid{0009-0000-0684-6742}
\par}
\cmsinstitute{Institut f\"{u}r Hochenergiephysik, Vienna, Austria}
{\tolerance=6000
W.~Adam\cmsorcid{0000-0001-9099-4341}, J.W.~Andrejkovic, T.~Bergauer\cmsorcid{0000-0002-5786-0293}, S.~Chatterjee\cmsorcid{0000-0003-2660-0349}, M.~Dragicevic\cmsorcid{0000-0003-1967-6783}, A.~Escalante~Del~Valle\cmsorcid{0000-0002-9702-6359}, R.~Fr\"{u}hwirth\cmsAuthorMark{2}\cmsorcid{0000-0002-0054-3369}, M.~Jeitler\cmsAuthorMark{2}\cmsorcid{0000-0002-5141-9560}, N.~Krammer\cmsorcid{0000-0002-0548-0985}, L.~Lechner\cmsorcid{0000-0002-3065-1141}, D.~Liko\cmsorcid{0000-0002-3380-473X}, I.~Mikulec\cmsorcid{0000-0003-0385-2746}, P.~Paulitsch, F.M.~Pitters, J.~Schieck\cmsAuthorMark{2}\cmsorcid{0000-0002-1058-8093}, R.~Sch\"{o}fbeck\cmsorcid{0000-0002-2332-8784}, D.~Schwarz\cmsorcid{0000-0002-3821-7331}, S.~Templ\cmsorcid{0000-0003-3137-5692}, W.~Waltenberger\cmsorcid{0000-0002-6215-7228}, C.-E.~Wulz\cmsAuthorMark{2}\cmsorcid{0000-0001-9226-5812}
% ...
\par}
\vskip\cmsinstskip
\dag:~Deceased\\
$^{1}$Also at Yerevan State University, Yerevan, Armenia\\
$^{2}$Also at TU Wien, Vienna, Austria\\
% ...
